# Manipulation of anisotropic Zhang-Rice exciton in van der Waals antiferromagnets NiPS$_{3-x}$Se$_x$ by anion substitution

Deepu Kumar[1], Joydev Khatua[2], Nguyen The Hoang[1], Yumin Sim[1], Rajesh Kumar Ulaganathan[3], Raju Kalaivanan[4], Raman Sankar[4], Maeng-Je Seong[1,*], and Kwang-Yong Choi[2,*]

[1]*Department of Physics and Center for Berry Curvature-based New Phenomena (BeCaP) Chung-Ang University, Seoul 06974, Republic of Korea*
[2]*Department of Physics, Sungkyunkwan University, Suwon 16419, Republic of Korea*
[3]*Centre for Nanotechnology, Indian Institute of Technology Roorkee, 247667, India*
[4]*Institute of Physics, Academia Sinica, Nankang, Taipei 11529, Taiwan*

**Abstract**

Spin-entangled excitons have emerged as intriguing quasi-particle excitations in van der Waals magnets. Among them, the recently observed Zhang-Rice (ZR) exciton in NiPS$_3$ has garnered significant research interest due to its strong correlation with magnetic ordering and its exceptionally long-lived coherence. Herein, we present our in-depth temperature- and polarization-dependent photoluminescence (PL) study of anion-substituted NiPS$_{3-x}$Se$_x$ ( $x = 0.008\ 0.03, 0.06$ and $0.09$ ) to explore the nature and dynamics of the ZR exciton. Our results reveal that, similar to the cation substitution[1,2], a small percentage of anion substitution effectively destroys and modulates the ZR exciton, as evidenced by the emergence of a weaker, lower-energy PL peak in addition to the primary ZR peak. The primary and secondary PL peaks exhibit the same anisotropic polarization but differ in their peak energy shift and intensity evolution with Se substitution, suggesting varying charge transfers of *p*-orbitals. Notably, the ZR exciton undergoes rapid thermal destabilization at much lower temperatures than two-magnon excitations, highlighting that *p*-orbital inhomogeneity beyond the magnetic ordering structure is a decisive factor in driving its thermal quenching.

\* Email: mseong@cau.ac.kr (M. J. Seong), choisky99@skku.edu (K.Y. Choi)

**Introduction**

A mutual coupling between various degrees of freedom such as spin, charge, lattice, and orbit gives rise to novel bound states, many-body phenomena, hybrid quasiparticle excitations, spin-orbit entangled excitons, and strong electron correlations[3–10]. Understanding the correlation between excitons and magnetic ordering in magnetic materials remains a fundamental challenge in the field of magneto-optics, spintronics, and quantum magnetism. In this direction, the spin-orbital entangled Zhang–Rice (ZR) state[11], has emerged as a promising excitonic state, highlighting the significant correlation between magnetism and electronic/optical degrees of freedom in dictating magneto-optical phenomena[12,13].

Among the various van der Waals (vdW) magnetic 2D materials, transition-metal thiophosphates $MPX_3$ have ushered in a new era of research in 2D magnetism due to their persistence of antiferromagnetism down to the atomically thin limit[14,15]. Within the $MPX_3$ family, $NiPS_3$, a self-doped negative charge-transfer antiferromagnetic insulator ($T_N$~150-155 K)[16,17] with an optical band gap of ~1.8 eV [9], offers a unique platform to explore the magneto-optical phenomena and the interplay between excitons and magnetic ordering. Recent studies have reported an extremely sharp and ultranarrow linewidth ($\sim 200-400\mu eV$) excitonic emission in $NiPS_3$ [8,18–20], which is extremely narrower than exciton emissions observed in other 2D magnetic materials[21] as well as non-magnetic transition metal dichalcogenides (e.g. $MoS_2$ and $WS_2$)[22]. This excitonic emission in $NiPS_3$ strongly couples with cavity photons, resulting in the formation of exciton-polaritons[11], and also couples with phonons, leading to the creation of exciton-phonon bound sidebands[8]. However, the origin of the exciton emission in $NiPS_3$ is still controversial, with different interpretations proposed, including spin-orbital entangled ZR excitons, defect-bound excitons, Hund excitons, and spin-flip induced *d-d* emission[6,7,13,23–25]. However, among these different scenarios, the ZR exciton arises from the transition between ZR triplet and ZR singlet states [13,18,20].

Beyond the parent compound NiPS$_3$, recent investigations on transition-metal site alloying have shown a rapid quenching of ZR exciton emission and a loss of its coherent nature with only a few percentage of substitution of either non-magnetic atoms (Ni$_{1-x}$Cd$_x$PS$_3$) or magnetic atoms (Ni$_{1-x}$Mn$_x$PS$_3$) at the Ni atom site, suggesting that cation substitution drastically destabilizes the ZR exciton[1,2]. Given this strong dependence on the cation substitution, an ensuing question arises: Can the ZR exciton also be influenced by chalcogen (anion) substitution? Understanding the role of anion substitution could provide deeper insights into the coherence and mechanism of excitonic and magneto-optical properties in NiPS$_3$-based systems.

To address this question and further investigate the impact of anion substitution on the ZR exciton, in this work, we perform temperature- and polarization-dependent PL measurements on Se-substituted NiPS$_{3-x}$Se$_x$ ( $x = 0.008\ 0.03, 0.06$ and $0.09$ ). For $x = 0.008$ Se-doped compound, we observe that the ZR exciton preserve its coherent nature and strong PL emission, similar to the parent NiPS$_3$. Surprisingly, with increasing Se content above $x = 0.03$, the ZR exciton is dramatically suppressed and significantly broadens, yet it survives up to our highest Se doping level of $x = 0.09$ studied. In contrast, magnetic ground state and magnetic excitations are not much affected by increasing Se concentration. This distinct impact of anion substitution on excitonic versus magnetic properties suggests that decoherence effects are highly sensitive to electronic heterogeneity.

**Results and Discussions**

**Characterization: Magnetic properties and magnon and phonon excitations**

Bulk NiPS$_3$ crystallizes in a monoclinic structure with space group $C2/m$ (#12) and point group $C_{2h}$. Figure 1a shows both the crystal and magnetic structures of NiPS$_3$, in which Ni atoms form a honeycomb lattice with zigzag chains of magnetic moments. Each Ni atom is surrounded by six S atoms, forming NiS$_6$ octahedra. Two P atoms, located above and below the Ni atom plane, are covalently bonded to six S atoms, forming a [P$_2$S$_6$]$^{4-}$ anion cluster. NiPS$_3$ undergoes an antiferromagnetic phase transition below $T_N \sim 150\text{-}158\text{ K}$, where the spins align ferromagnetically along zigzag chains parallel to the crystallographic *a*-axis with a slight out-of-plane component due to weak magnetic anisotropy along the *b*-axis[16,17,26–28]. These ferromagnetic zigzag chains are coupled antiferromagnetically within the *ab* plane, see Figure 1a. NiPSe$_3$ shares a similar magnetic structure as NiPS$_3$, showing magnetic ordering below $T_N \sim 212\text{ K}$ [17].

To verify phase purity, single-crystal X-ray diffraction (XRD) patterns of NiPS$_{3-x}$Se$_x$ were collected using a Bruker-AX (D8-Advance) X-ray diffractometer with Cu Kα radiation (λ = 1.54Å) at room temperature. Figure 1b presents the XRD data collected with the X-ray beam perpendicular to the (00l) planes, revealing the characteristic parallel planes of the monoclinic crystal structure. With increasing Se concentration at the S site, a systematic low-angle shift of the (00l) peaks is observed, see right panel of Fig. 1b, indicating a slight elongation of the *c*-axis due to the larger atomic radius of Se compared to S. This trend is consistent with previous reports[17]. We also measured the temperature-dependent magnetic susceptibility of NiPS$_{3-x}$Se$_x$, as shown in Figure 1c with an external magnetic field of $\mu_0 H = 1\text{T}$ applied along the *ab* plane. For all samples, we obtained the Néel temperature around $\sim 156\text{ K}$, from the temperature-dependent derivative of the magnetic susceptibility, see inset in Fig. 1b, which show good agreement with previous studies[16,17,27,28].

We next turn to Raman characterizations to examine magnetic and lattice excitations. Figure 1c illustrates the unpolarized Raman spectra of NiPS$_{3-x}$Se$_x$ collected at $T$=3.5 K. We observe five $B_g$ and three $A_g$ peaks in the frequency range of 100 to 650 cm$^{-1}$, which are labeled as $B_g^k$ ($k=1-5$) and $A_g^i$ ($i=1-5$), respectively, and are in line with one-phonon modes of parent NiPS$_3$[15,29,30], see Supporting Information for details on the expected phonons at the Brillouin zone center and polarization selection rules. We note that the $B_g^k$ ($k=1-5$) phonons are generally a combination of the two nearly degenerate phonon modes with the same frequency but different symmetries ($A_g$ and $B_g$) and have also been labeled as $A_g$ and $B_g$ in earlier Raman studies on NiPS$_3$. Here, we adopt the $B_g^k$ ($k=1-5$) notation for consistency. The $B_g^1$ and $B_g^2$ phonon modes involve the vibrations of the Ni$^{2+}$ ions, while the higher-frequency phonons above ~200 cm$^{-1}$ are expected to be associated with the vibrations of the $[P_2S_6]^{4-}$ cluster. Furthermore, we observed two additional phonons labeled as $P_1$ and $P_2$, which are likely related to the vibrations of the $[P_2Se_6]^{4-}$ cluster in NiPSe$_3$. Additionally, a broad and pronounced continuum centered around ~530 cm$^{-1}$ is attributed to the two-magnon (2$M$) excitation, which arises from double-spin-flip process with equal but opposite momenta [15].

**Robustness of two-magnon excitations against anion substitution**

Before discussing the dynamics of the ZR excitonic emission, we first address the Se substitution effect on the 2$M$ behavior and its coupling with phonons. Figure 2a-c presents the unpolarized color contour maps of the Raman intensity versus Raman shift across a temperature range of 3.5 – 330 K for the $x$ = 0.03, 0.06 and 0.09 samples. The sharp lines correspond to the phonon excitations, while the broad continuum corresponds to the 2$M$ excitation. As the temperature increases above ~100 K, the 2$M$ excitation softens and broadens. We carefully extracted the self-energy parameters of 2$M$ such as frequency and linewidth up to ~ 200 K using the Lorentzian function after subtracting phonon excitations. Figure 2d and Figure S2 summarize the temperature dependence of the 2$M$'s frequency, linewidth, and intensity. We observe no appreciable $x$ dependence, suggesting that the 2M excitations remain robust under Se substitution, at least up to $x$ ~ 0.09. This is fully consistent with the absence of variation in $T_N$ up to $x$ = 0.09.

To further track the 2M excitations up to our highest recorded temperature, we examine the evolution Fano asymmetry of the $B_g^2$ ~ 180 cm$^{-1}$ and $B_g^5$ ~ 560 cm$^{-1}$ phonons, see Figure S3. The asymmetric parameter is extracted using the Breit–Wigner Fano function[31] $I_{BWF}(\omega) \propto (1+\delta/q)^2 / 1+\delta^2$, where $\delta = \omega - \omega_0 / \Gamma$ and $\Gamma$ and $\omega_0$ are the linewidth and frequency of the uncoupled phonon, respectively. The asymmetric parameter $1/|q|$ quantifies the coupling strength between the phonon and electronic/magnetic continuum excitations. In the strong-coupling limit $1/|q| \to \infty$, the phonon line shape becomes increasingly asymmetric. As shown in Figure 2f, $1/|q|$ of the $B_g^5$ mode, located on the higher-energy side of the 2M continuum, shows a gradual decrease above ~100 K. This thermal evolution resembles the temperature dependence of the 2M peak frequency. Conversely, $1/|q|$ of the $B_g^2$ mode, which lies on the lower-energy side of the 2M continuum, starts to increase for temperatures above $T_N$, as shown in Figure 2e. This behavior suggests that the $B_g^2$ Fano phonon reflects the thermal damping of magnons into paramagnons as the system transitions through $T_N$. Noteworthy is that the Fano parameters of both $B_g^5$ and $B_g^2$ modes exhibit little variation with $x$, confirming the robustness of the 2M signals against anion substitution up to the concentration of $x\sim0.09$.

**Anion-substitution and temperature dependence of PL emission**

Next, we turn to the effect of anion substitution on the dynamics of the ZR exciton. To check the homogeneity of the samples, we collected PL spectra at $T = 3.5$ K (see Figure S6) from distinct spots that cover several micrometers across the sample. The consistent PL emission observed from all spots excludes the presence of micrometer-sized inhomogeneities. Figure 3a depicts the PL spectrum of the NiPS$_{3-x}$Se$_x$ collected at 3.5 K using $\lambda=$ 532 nm (2.33 eV) laser excitation. The PL spectrum for $x = 0.008$ shows a sharp and ultra-narrow PL emission (labeled as $X_{ZR_2}$) located at $\sim1.47543$ eV, with a linewidth of $\sim389.2$ μeV. This PL peak can be attributed to the main ZR excitonic emission[1,2,13]. In addition to the primary ZR excitonic peak, a very weak peak labeled as $X_{ZR_1}$ is observed at $\sim1.47309$ eV and is found to be an order of magnitude broader ($\sim1.82791$ meV) compared to the $X_{ZR_2}$ peak, see Table I for detailed peak positions and linewidths/full width at half maximum (FHWM).

A striking difference between these excitons is their opposite energy shifts with Se substitution. The $X_{ZR_2}$ ($X_{ZR_1}$) peak shows a blue (red) shift with increasing $x$, which leads to an increase in energy separation between these two excitonic features, see Figure 3b,c. This intriguing opposite trend is further confirmed by PL spectra collected from different spots of the sample (see Figure S7), ruling out chemical inhomogeneities as its origin. Given the small energy difference between $X_{ZR_1}$ and $X_{ZR_2}$, the secondary peak cannot be ascribed to an exciton-phonon sideband. Rather, it appears to be an intrinsic property of the material, possibly arising from the presence of two slightly different electronic and magnetic states generated by Se substitution. As the Se concentration increases from $x=0.008$ to $x=0.03$, the main $X_{ZR_2}$ peak is drastically suppressed, followed by a gradual decrease in intensity with further substitution up to $x=0.09$ (see Figure 3a,e). In contrast to $X_{ZR_2}$, $X_{ZR_1}$ is relatively less affected by anion substitution, showing a gradual intensity decrease with increasing $x$, see Figure 3e. On the other hand, both $X_{ZR_1}$ and $X_{ZR_2}$ undergo a quasi-linear broadening with increasing Se concentration see Figure 3d.

Next, we examine the thermal effect of the ZR exciton. Figure 4a-d shows the temperature-dependent PL color maps for NiPS$_{3-x}$Se$_x$ (see Figure S8a-d for the temperature evolution of the PL spectra). For $x=0.008$, the intensity of the $X_{ZR_2}$ remains nearly temperature-independent up to 40 K and then gradually decreases with further temperature rise and completely vanishes above $T^*\sim 120$ K, which is $\sim 30$ K lower than $T_N$. The linewidth of the $X_{ZR_2}$ increases with temperature, accompanied by a red shift in its peak energy. Furthermore, the vanishing temperature of the $X_{ZR_2}$ exciton decreases progressively with increasing Se concentration: $T^*\sim 110\text{-}100$ K for $x=0.03$, $T^*\sim 100$ K for $x=0.06$, and $T^*\sim 90$ K for $x=0.09$. $X_{ZR_1}$ shows a similar temperature dependence as $X_{ZR_2}$, but it vanishes at a lower temperature than the primary peak. This disappearance of the ZR exciton at temperatures much below $T_N$ reflects that the ZR exciton is related to the magnetic ordering, but additional mechanisms are involved in the destabilization of the ZR exciton at much lower temperatures than $T_N$, which will be discussed later.

To gain quantitative insight into the thermal effect on the ZR exciton, we plot the peak energy and linewidth as a function of temperature for $x=0.008$ in Figure 4e,f, respectively.

Generally, the temperature dependence of excitonic emission energy in semiconductors is described using the empirical relation suggested by Donnell and Chen[32], which accounts for electron-phonon interactions in exciton energy shifts with temperature:

$$E(T) = E_0 - SE_p [\coth(\frac{E_p}{2k_B T}) - 1] \quad (1)$$

where $E_0$ is the exciton energy at $T$=0 K, $E_p = \langle \hbar\omega \rangle$ is the average phonon energy contributing to the shift in exciton energy, $k_B$ is the Boltzmann constant, and $S$ is the dimensionless Huang–Rhys factor describing the strength of exciton-phonon coupling. The solid red lines in Figure 4e are the fitting curve using equation (1), with the best-fit parameters listed in Table II. Moreover, the temperature-dependent linewidth of the exciton can also be described by considering exciton-phonon interactions. At finite temperatures, the temperature-dependent linewidth of the exciton is given as[33]

$$\Gamma(T) = \Gamma_0 + \frac{\lambda_{LO}}{(\exp(\Theta_{LO}/T) - 1)} \quad (2)$$

Where $\Gamma_0$ is the exciton linewidth at $T$=0 K, which arises from scattering of excitons with impurities and imperfections. The second term is associated with the linewidth broadening due to the exciton scattering with longitudinal optical ($LO$) phonons, where $\lambda_{LO}$ represents the exciton-LO phonon coupling strength, and $\Theta_{LO}$ denotes the LO phonon energy. The solid red lines in Figure 4f are the fitted curves using equation (2), and the fit parameters are listed in Table II. The aforementioned models as given in equations (1) and (2) are in good agreement with the experimental data, reflecting that the exciton-phonon mechanism seems to provide a consistent description of the observed energy redshift and linewidth broadening with increasing temperature. Nonetheless, it is far from clear whether the exciton-phonon mechanism alone can fully account for the thermal behavior of the ZR exciton, particularly given the inability to obtain reliable parameters for $x = 0.03 - 0.09$.

**Linearly polarization-dependent PL**

As aforementioned, the ferromagnetic zigzag spin chains along with the Néel vector align along the $a$-axis. It is extremely important to figure out the correlation between the ZR excitonic emission and the spin structure in the chalcogen-substituted NiPS$_3$. An intriguing characteristic of the ZR excitonic emission in NiPS$_3$ is that it shows a maximum intensity when the collection polarization is perpendicular to the crystallographic $a$-axis (Néel vector), indicating a highly anisotropic linear polarization of the ZR excitonic emission[19]. To explore this correlation in Se-mixed NiPS$_3$, we conducted linear polarization-dependent PL measurements on the $x = 0.008$ and $x = 0.03$ samples in a parallel configuration. Simultaneously, we also performed linearly polarized Raman measurements on the same sample spot to determine the crystal symmetry and the crystallographic axis. For the higher-$x$ samples, the limited statistics hinder analysis of the detailed angular dependence.

Figure 5 a sketches the schematic representation of our experimental setup for linearly polarized Raman and PL measurements. Figure 5 b,c shows the intensity polar plots for Raman-active phonons $B_g^1$ and $B_g^2$ for $x = 0.03$ collected at 3.5 K in the parallel configuration ($\hat{e}_i \parallel \hat{e}_s$) after correcting the initial orientation of the crystal axis, see supplementary information for the Raman selection rules and as-measured uncorrected polarized Raman spectra, as well as intensity polar plots of the phonon excitations in Figure S4 for $x = 0.03$ and Figure S5 for $x = 0.008$. The observed splitting of $B_g^2$ into two components is not surprising, as such splitting is expected in the magnetically ordered phase due to the breaking of the three-fold rotation symmetry of the lattice[15]. Both split components show a four-lobed symmetry, which are in good agreement with the earlier linearly polarized Raman studies [34]. Figure 5d shows the PL spectra at 3.5K for two configurations, i.e., parallel ($\hat{e}_i \parallel \hat{e}_s$) and perpendicular ($\hat{e}_i \perp \hat{e}_s$) configurations. We observe the highly anisotropic excitonic emission with a nearly unity degree of linear polarization, implying a strong coupling between the ZR exciton and zig-zag spin chain structure[34]. Next, we plot the intensity polar plot for $X_{ZR_1}$ and $X_{ZR_2}$ as a function of polarization angle in Figure 5e,f, measured in the parallel configuration, after correcting the orientation of the crystal axis according to the intensity polar plot of the $B_g^1$ and $B_g^2$ modes, see Figure S9b for the uncorrected polarization-angle-dependent PL spectrum. The linear polarization results for $x=0.008$ are shown in Figure S9a and Figure S10. For $x=0.008$ and $x=0.03$, both of the $X_{ZR_1}$ and $X_{ZR_2}$ excitonic emissions show the maximum (minimum) intensity along the *b*-axis (*a*-axis/Néel vector), being in line with what was observed in parent NiPS$_3$[19,34].

**Discussion**

We have investigated the dynamics of the ZR exciton by combining temperature- and linearly polarization-dependent PL measurements for NiPS$_{3-x}$Se$_x$. For $x=0.008$, the primary ZR exciton is found to have similar characteristics (ultra-narrow linewidth and intense PL emission) as observed in the parent NiPS$_3$. Surprisingly, with increasing Se concentration, the ZR exciton undergoes a rapid quenching in its PL intensity and significant linewidth broadening above $x=0.03$. This drastic disruption of the ZR exciton induced by anion substitution closely resembles the behavior reported in cation-substituted compounds Ni$_{1-x}$Cd$_x$PS$_3$ or Ni$_{1-x}$Mn$_x$PS$_3$[1,2]. Our findings reveal that even a small percentage of heterogeneous metal or anion substitutions are detrimental to the ZR excitons. Nonetheless, the decoherence mechanism may differ. For example, in the case of magnetic metal-ion substitution, a double-spin-flip process plays a significant role in the thermal behavior of ZR excitons.

In contrast to the ZR exciton, the magnetic ground state and $2M$ excitations are not much affected by the anion substitution in the investigated range of $x = 0.008$-$0.09$. Furthermore, temperature-dependent measurements reveal that the 2M excitations remain robust up to $\sim 0.5\ T_N$ irrespective of $x$, while the ZR exciton becomes destabilized at much lower temperatures of $\sim 0.25\ T_N$, reflecting that the magnetic ground state and magnon dynamics are not sufficient to explain the rapid quenching of the ZR exciton with anion substitution or the observed thermal effects. Additionally, the main ZR shows a blueshift in its peak energy with an increase in Se concentration, which contradicts the expected band gap trend, as the band gap is advocated to decrease from $NiPS_3$ to $NiPSe_3$[35,36]. This unexpected increase in the primary ZR energy with increasing Se concentration is not consistent with changes in the band gap. On the other hand, a second PL peak appears on the lower energy side of the main ZR PL peak and exhibits the identical anisotropic polarization characteristic as the main ZR peak. However, in contrast to the main ZR peak, this low-energy peak shows the following differences: (i) It is very weak and is an order of magnitude broader for $x = 0.008$, but it becomes equally intense as the main ZR peak for $x \geq 0.03$. (ii) It shows a red-shift trend with increase in Se concentration. (iii) This peak vanishes at a temperature much lower than the main ZR peak's destruction temperature. Since the ZR exciton originates from the spin and charge entanglement between the Ni $d$ orbital and the ligand (S/Se) $p$ orbital, the appearance of two distinct PL peaks upon Se substitution can be attributed to variations in charge transfer between the $p$ orbitals of S and Se, thereby engendering the self-organization of two different spin- and charge-entangled states.

In summary, our comprehensive study of the temperature, composition, and angle-resolved PL in NiPS$_{3-x}$Se$_x$ ($x$ = 0.008-0.09) revealed that even a slight degree of electronic heterogeneity in the $p$ orbitals rapidly destabilizes the primary spin-entangled exciton, leading to the emergence of a secondary exciton. These findings highlight the decisive role of electronic uniformity in stabilizing entangled magneto-excitons.

**Acknowledgments:**

This work was supported by the National Research Foundation (NRF) of Korea (Grant Nos. 2020R1A5A1016518, 2022R1A2C1003959, and RS-2023-00209121). R.S. acknowledges the financial support provided by the Ministry of Science and Technology in Taiwan under Project No. NSTC (113-2124-M-001-045- MY3 and 113-2124-M-001-003), Financial support from the Center of Atomic Initiative for New Materials (AI-Mat), National Taiwan University, (Project No. 113L900801) and Academia Sinica for the budget of AS- iMATE-113-12. R.K.U. would like to acknowledge the IITR for the Faculty Initiation Grant (FIG-101068).

**Table I:** Peak energy and FWHM of the $X_{ZR_1}$ and $X_{ZR_2}$ excitonic emission for NiPS$_{3-x}$Se$_x$.

| NiPS$_{3-x}$Se$_x$ | $X_{ZR_1}$ | | $X_{ZR_2}$ | |
|---|---|---|---|---|
| | Energy [eV] | FWHM [meV] | Energy [eV] | FWHM [meV] |
| x=0.008 | 1.47309 | 1.82791 | 1.47543 | 0.38921 |
| x=0.03 | 1.47252 | 2.82077 | 1.47584 | 1.16396 |
| x=0.06 | 1.47242 | 3.25464 | 1.47616 | 1.60424 |
| x=0.09 | 1.47201 | 4.56365 | 1.47625 | 2.63723 |

**Table II:** List of fitting parameters obtained from the fitting of temperature-dependent exciton energy and linewidth for $x = 0.008$ as described in the text.

| | $E_0$ [eV] | $S$ | $E_p$ [meV] | $\Gamma_0$ [meV] | $\lambda_{LO}$ [meV] | $\Theta_{LO}$ [K] |
|---|---|---|---|---|---|---|
| $X_{ZR_1}$ | 1.47312±3.44E-5 | 1.29±0.4 | 27.95±2.7 | 1.82±0.05 | 2.64±0.9 | 60 ±12 |
| $X_{ZR_2}$ | 1.47542±2.76E-6 | 0.84±0.06 | 20.5±0.6 | 0.397±0.006 | 38.37±3.9 | 196±6 |

**Figure:**

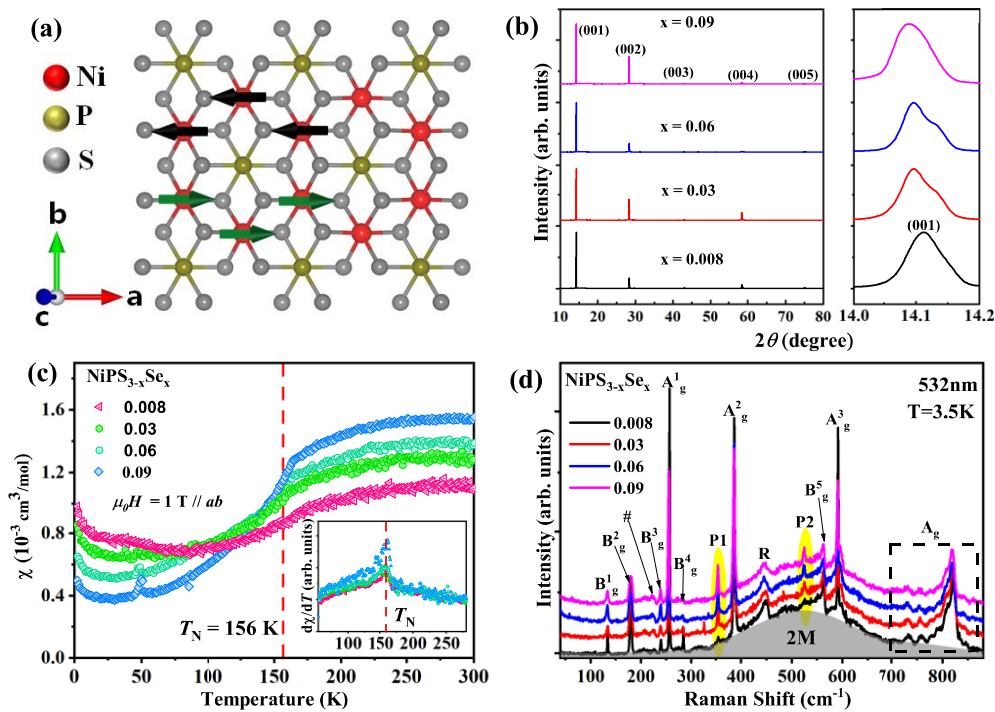

**Figure 1: (a)** Crystal and magnetic structure of NiPS$_3$. Red and green arrows indicate the spin orientations of Ni atoms in the magnetically ordered phase. **(b)** XRD patterns for the NiPS$_{3-x}$Se$_x$ series. **(c)** Temperature dependence of the magnetic susceptibility for NiPS$_{3-x}$Se$_x$ along the $ab$ plane measured under an external field of $\mu_0 H = 1\,\text{T}$. The inset in (c) shows the temperature-dependent derivative of the magnetic susceptibility. The red dashed line in (c) and the inset marks the Néel temperature ($T_N \sim 156\,\text{K}$). **(d)** Unpolarized Raman spectra of NiPS$_{3-x}$Se$_x$, collected at $3.5\,\text{K}$ using $532\,\text{nm}$ ($2.33\,\text{eV}$) excitation. The spectra are vertically shifted for clarity, see Figure S1 for spectra plotted on the same scale. The observed phonon excitations in the spectral range of $100\text{-}650\,\text{cm}^{-1}$ are labeled as $B_g^{1-5}$, $A_g^{1-3}$, R and P1-P2. The gray-shaded area corresponds to the two-magnon ($2M$) continuum.

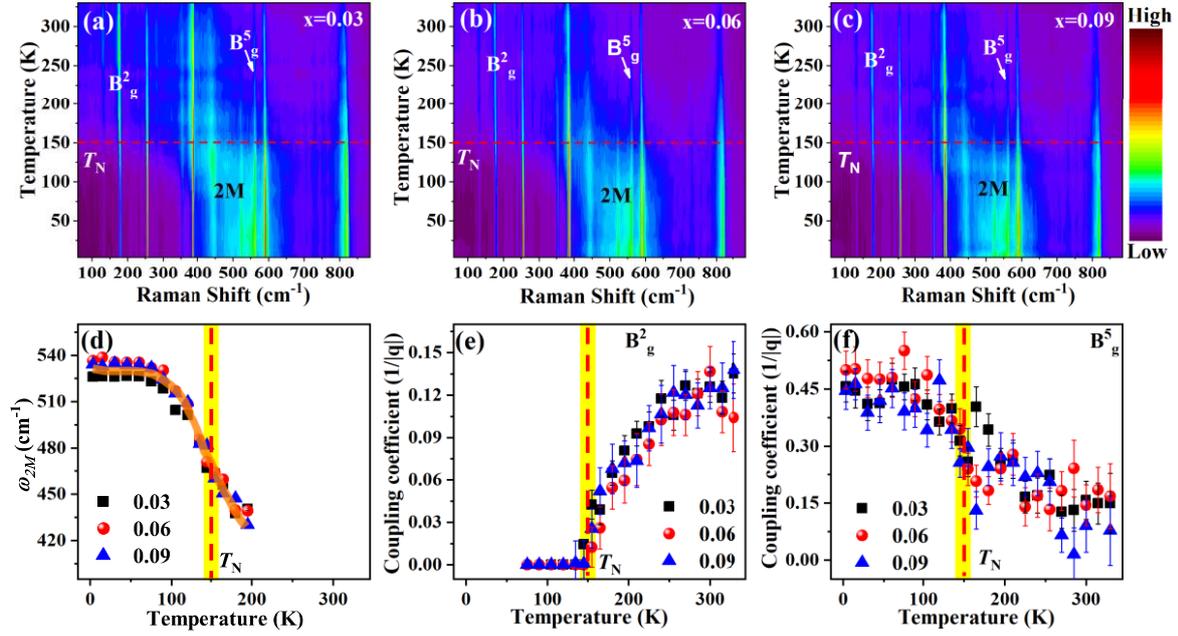

**Figure 2:** Unpolarized 2D color contour maps of Raman intensity versus Raman shift and as a function of temperature for **(a)** $x= 0.03$ **(b)** 0.06 and **(c)** 0.09 of NiPS$_{3-x}$Se$_x$. Sharp lines represent phonon excitations, while the broad continuum represents the $2M$ excitation. **(d)** Frequency of $2M$ as a function of temperature. **(e)** and **(f)** Fano resonance coupling coefficient $1/|q|$ of the $B_g^2$ and $B_g^5$ mode as a function of temperature. The red horizontal lines in **(a-c)** and vertical dashed lines in **(d-f)** denote the antiferromagnetic phase transition temperature ($T_N$).

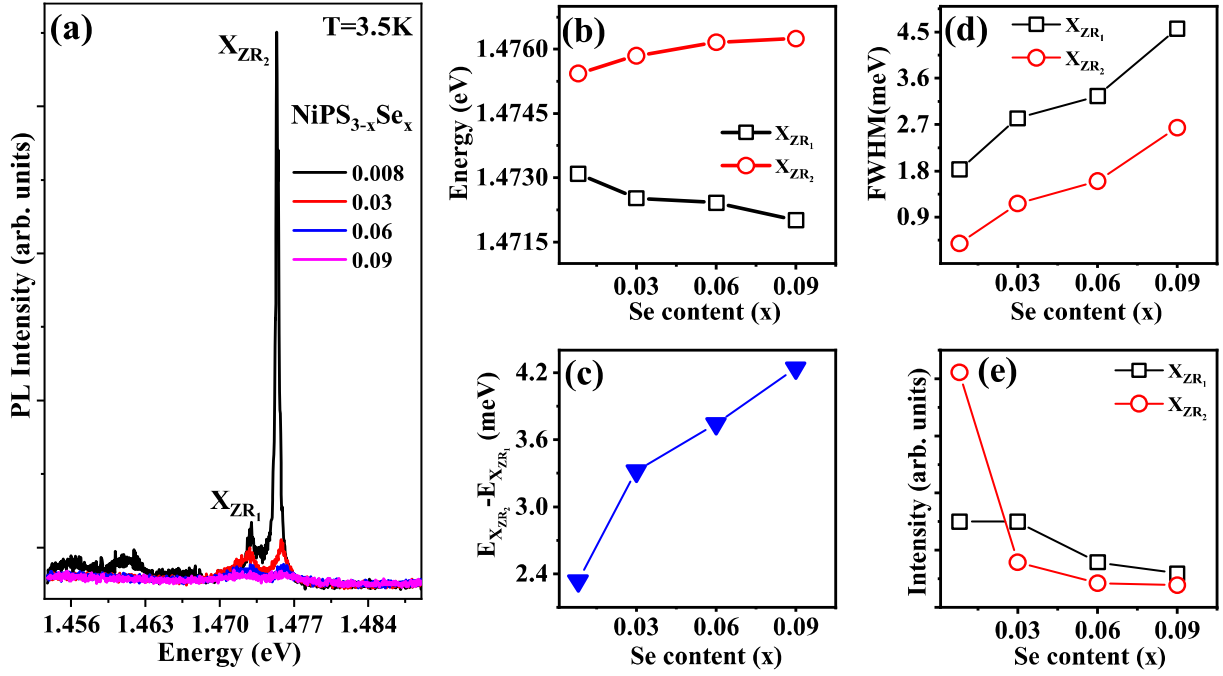

**Figure 3: (a)** Unpolarized PL spectra of NiPS$_{3-x}$Se$_x$, collected at 3.5 K as a function of Se concentration using 532 nm (2.33 eV) excitation. $X_{ZR_2}$ and $X_{ZR_1}$ correspond to the primary and secondary ZR PL emissions, respectively. **(b)** Peak energy of $X_{ZR_2}$ and $X_{ZR_1}$ versus Se content. **(c)** Energy difference between the $X_{ZR_2}$ and $X_{ZR_1}$ peaks versus Se content. **(d)** and **(e)** FWHM and integrated intensity of $X_{ZR_2}$ and $X_{ZR_1}$ as a function of Se content, respectively.

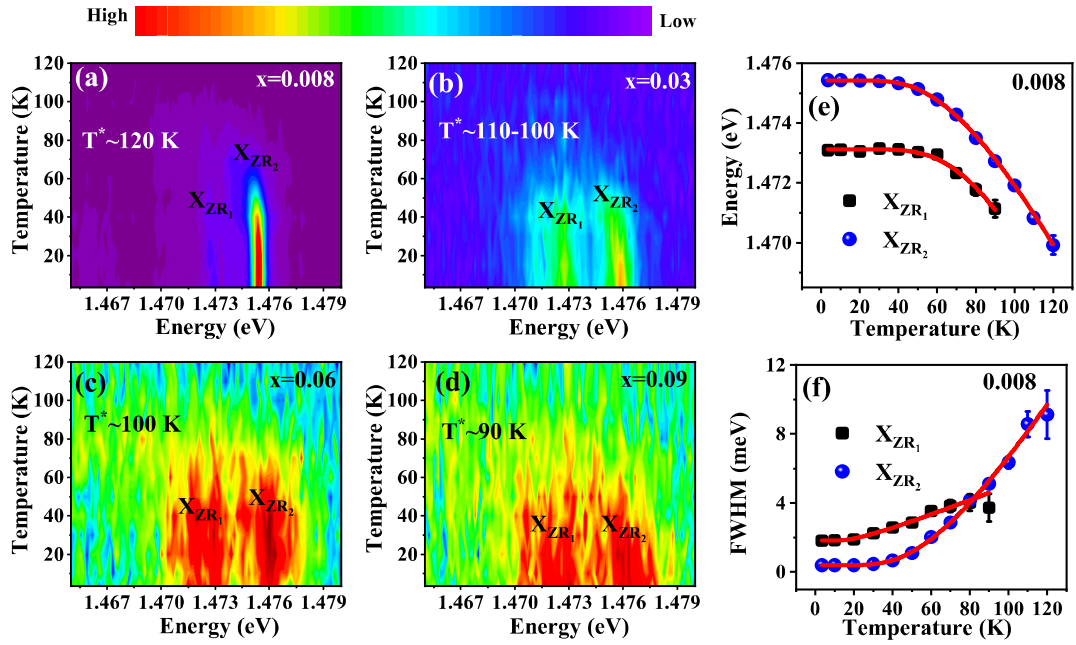

**Figure 4: (a-d)** Unpolarized 2D color contour maps of PL intensity versus PL energy and as a function of temperature for NiPS$_{3-x}$Se$_x$. $T^*$ is the temperature, above which $X_{ZR_2}$ become too weak to detect. **(e)** and **(f)** Temperature-dependent peak energy and FWHM of $X_{ZR_2}$ and $X_{ZR_1}$ for $x = 0.008$, respectively. The solid lines in **(e)** and **(f)** represent the fitted curves as described in the text.

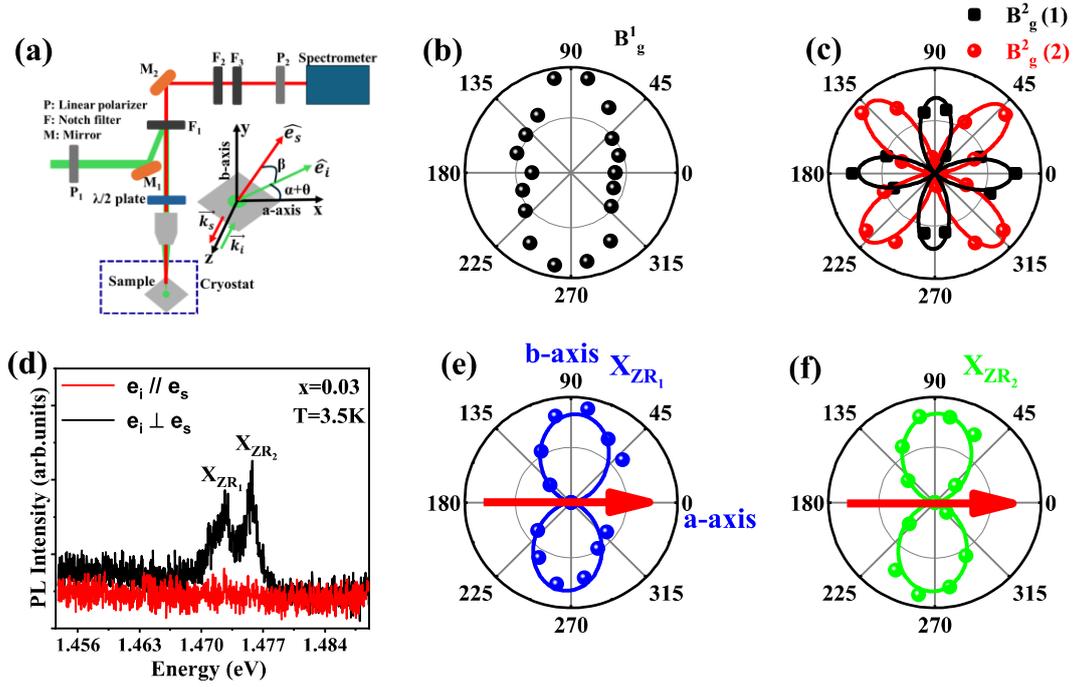

**Figure 5: (a)** Schematic representation of the experimental setup for linearly polarized Raman and PL measurements with polarization directions of incident and scattered light. **(b)** and **(c)** Intensity polar plot for the Raman phonons $B_g^1$ and $B_g^2(1,2)$ for $x=0.03$, collected at 3.5 K in the parallel configuration ($\hat{e}_i \parallel \hat{e}_s$), after the corrections of the initial orientation of the crystal axis, respectively. **(d)** PL spectra for $\hat{e}_i \parallel \hat{e}_s$ (red) and $\hat{e}_i \perp \hat{e}_s$ (black) for $x=0.03$ at 3.5 K. **(e)** and **(f)** Intensity polar plot of the excitonic peaks $X_{ZR_2}$ and $X_{ZR_1}$ for $x=0.03$ in the parallel configuration at 3.5 K, after correcting the initial crystal axis orientation, respectively. The red arrows in **(f)** and **(e)** represent the Neel vector, which aligns along the *a*-axis. The angular dependence of the $B_g^2(1)$ an $B_g^2(2)$ intensities are described by $I(\theta)=e^2\cos^2(2\theta)$ and $I(\theta)=e^2\sin^2(2\theta)$, respectively, while the excitonic peaks $X_{ZR_2}$ and $X_{ZR_1}$ follow $I(\theta)=I_0\sin^2(\theta)$.

**Supporting Information:**

**Manipulation of anisotropic Zhang-Rice exciton in van der Waals antiferromagnets NiPS$_{3-x}$Se$_x$ by anion substitution**

Deepu Kumar[1], Joydeep Khatua[2], Nguyen The Hoang[1], Yumin Sim[1], Rajesh Kumar Ulaganathan[3], Raju Kalaivanan[4], Raman Sankar[4], Maeng-Je Seong[1,*], and Kwang-Yong Choi[2,*]

[1]*Department of Physics and Center for Berry Curvature-based New Phenomena (BeCaP) Chung-Ang University, Seoul 06974, Republic of Korea*
[2]*Department of Physics, Sungkyunkwan University, Suwon 16419, Republic of Korea*
[3]*Centre for Nanotechnology, Indian Institute of Technology Roorkee, 247667, India*
[4]*Institute of Physics, Academia Sinica, Nankang, Taipei 11529, Taiwan*

**Sample growth:**

High-quality NiPS$_{3-x}$Se$_x$ (x = 0.008, 0.03, 0.06, and 0.09) single crystals were grown using the chemical vapor transport (CVT) method with iodine as the transport agent. Initially, polycrystalline powders were synthesized via solid-state reaction under high-vacuum conditions. High-purity starting materials—nickel powder (99.999%), phosphorus powder (99.999%), sulfur powder (99.999%), and selenium powder (99.999%)—were weighed in the nominal stoichiometric ratio and sealed in a quartz tube (22 mm in diameter) under a vacuum of $10^{-3}$ Torr. The mixture was then subjected to two-stage heating at 400 °C and 600 °C, with intermittent grinding, to ensure the formation of a single-phase compound. For crystal growth, 200 mg of iodine was added to the synthesized polycrystalline powder, which was then sealed in a quartz tube (dimensions: 20 mm × 22 mm × 400 mm) under a vacuum of $10^{-3}$ Torr. The tube was placed in a two-zone furnace, with temperature zones set at 700 °C and 600 °C, and maintained for 200 hours. After the growth process, the furnace was cooled to room temperature at a rate of 2 °C/min. The quartz tube was subsequently broken inside an argon-filled glovebox, and high-quality single crystals were collected.

**Temperature-and polarization-dependent Raman and Photoluminescence measurements**

Raman and photoluminescence spectroscopic measurements were done using the Princeton SpectraPro HRS-750 spectrometer in backscattering geometry. A 532 nm (2.33 eV) laser excitation source was used to excite the Raman and photoluminescence spectra. Laser power on the sample was kept below ~500 $\mu$W to avoid any local heating and to prevent sample damage. A 50x objective lens was used to focus the incident light beam onto the sample and the same objective lens was used to collect the scattered light beam from the sample. The scattered light was dispersed using a 1200 grating coupled with an electrically cooled BLAZE charge coupled device detector. Temperature-dependent Raman and photoluminescence measurements were carried out using a closed-cycle cryostat (Montana Cryostat) by varying the temperature from 3.5 to 330 K, under high vacuum with a temperature accuracy of ± 0.1 K.

To ensure thermal equilibrium, a waiting time of ~15 minutes was maintained before acquiring each Raman/PL spectrum. A set of linear polarizers and a half-wave plate were used to perform polarization-dependent PL measurements. A vertical analyzer/polarizer was installed in front of the spectrometer to keep the consistent signal with respect to the grating orientation (see Figure 5, in the main text for a schematic representation). A half-wave plate was installed to rotate the polarization angle in the parallel configuration.

Temperature-dependent magnetic susceptibility measurements were performed using a superconducting quantum interference device vibrating sample magnetometer (SQUID VSM, Quantum Design) with an applied magnetic field of $\mu_0 H = 1\,\text{T}$ along the *ab* plane.

**Phonon excitation and selection rules**

The unit cell of bulk NiPS$_3$ gives rise to 30 phonon excitations at the $\Gamma$ point with the following irreducible representations: $\Gamma = 8A_g + 7B_g + 6A_u + 9B_u$, with 15 Raman-active representations ($\Gamma = 8A_g + 7B_g$), 12 infrared-active ($\Gamma = 5A_u + 7B_u$), and the remaining 3 ($\Gamma = A_u + 2B_u$) acoustic phonons[1].

The Raman tensors for Raman-active $A_g$ and $B_g$ phonons with $C_{2h}$ point symmetry group can be expressed as[2,3]

$$R(A_g) = \begin{pmatrix} a & 0 & d \\ 0 & b & 0 \\ d & 0 & c \end{pmatrix} \text{ and } R(B_g) = \begin{pmatrix} 0 & e & 0 \\ e & 0 & f \\ 0 & f & 0 \end{pmatrix}.$$

Within the semi-classical approximation, the Raman scattering intensity of the first-order phonon modes is given as $I_{\text{int}} = |\hat{e}_s^t R \hat{e}_i|^2$, where $\hat{e}_i$ and $\hat{e}_s$ are the polarization direction unit vector of the incident and scattered light, respectively[2,3].

As both the incidents and scattered light are polarized on the XY or *ab* plane, the polarization direction unit vectors $\hat{e}_i$ and $\hat{e}_s$ may be decomposed as $\hat{e}_i = [\cos(\theta+\alpha), \sin(\theta+\alpha), 0]$ and $\hat{e}_s = [\cos(\theta+\alpha+\beta), \sin(\theta+\alpha+\beta), 0]$, respectively, where $\theta$ is the polarization angle measured relative to the crystallographic *a*-axis. $\beta$ is the angle between the incident and scattered light polarization direction and for the parallel- and cross-configurations, $\beta$ corresponds to $0°$ and $90°$, respectively. The constant angle $\alpha$ arises from some deviation in the incident light polarization direction from the crystallographic *a*-axis and if the incident light polarization direction is perfectly aligned along the *a*-axis, $\alpha$ can be considered as 0. For the parallel configuration ($\hat{e}_i = \hat{e}_s = [\cos(\theta+\alpha), \sin(\theta+\alpha), 0]$), The Raman scattering intensity of the $A_g$ and $B_g$ mode with $C_{2h}$ point group is given as $I_{A_g} = |a\cos^2(\theta+\alpha) + b\sin^2(\theta+\alpha)|^2$ and $I_{B_g} = |e\sin 2(\theta+\alpha)|^2$. The degree of linear polarization for PL emission is defined as

$$\rho = I_b - I_a / I_b + I_a$$

$I_b (I_a)$ is the intensity of the PL emission when the polarization direction is along the *b*-axis (*a*-axis).

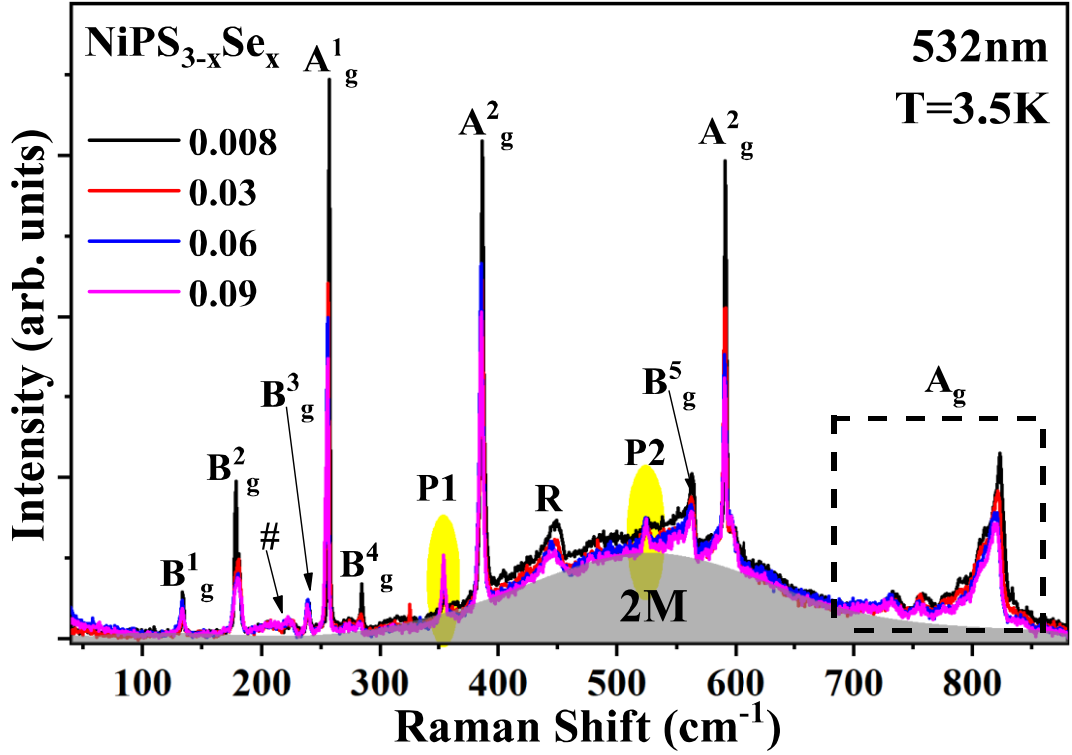

**Figure S1:** Unpolarized Raman spectra of NiPS$_{3-x}$Se$_x$, collected at 3.5 K as a function of Se concentration using 532 nm (2.33 eV) laser excitation.

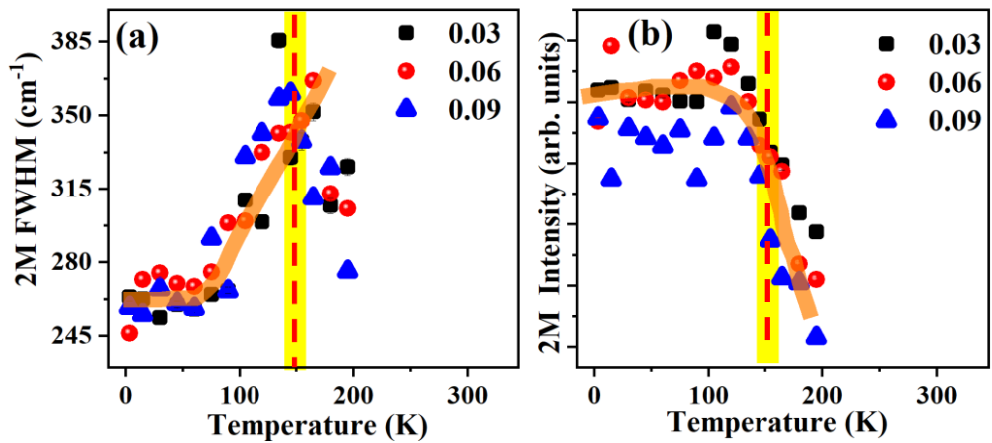

**Figure S2:** Temperature-dependent **(a)** FWHM and **(b)** intensity of 2$M$ excitation.

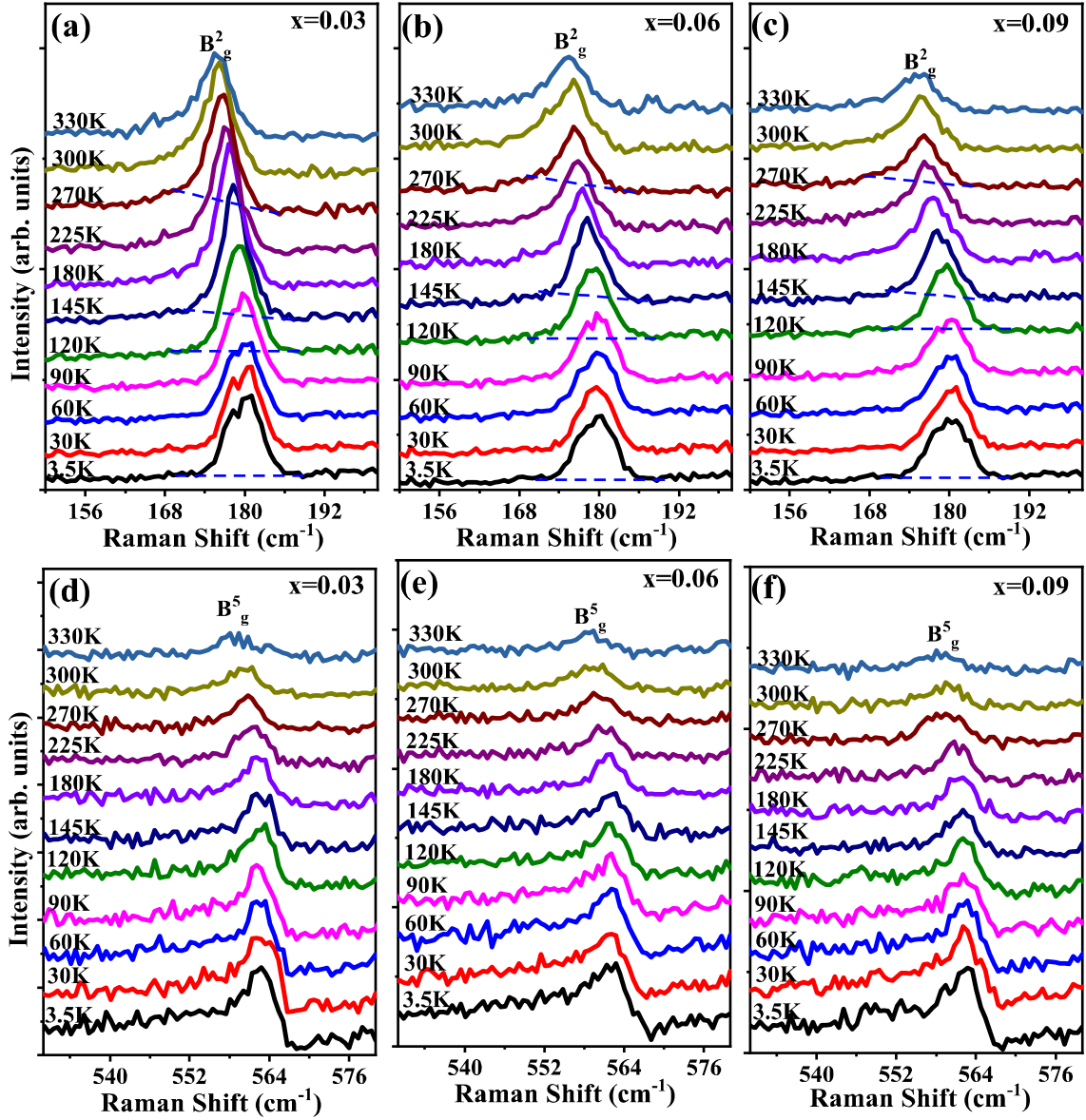

**Figure S3:** **(a-c)** and **(d-f)** Temperature evolution of the unpolarized Raman spectrum of NiPS$_{3-x}$Se$_x$ in the frequency range of $150-200$ cm$^{-1}$ and $530-580$ cm$^{-1}$, respectively. Dashed blue lines in (a-c) represent the evolution of the asymmetry for the $B_g^2$ mode with the rise in temperature.

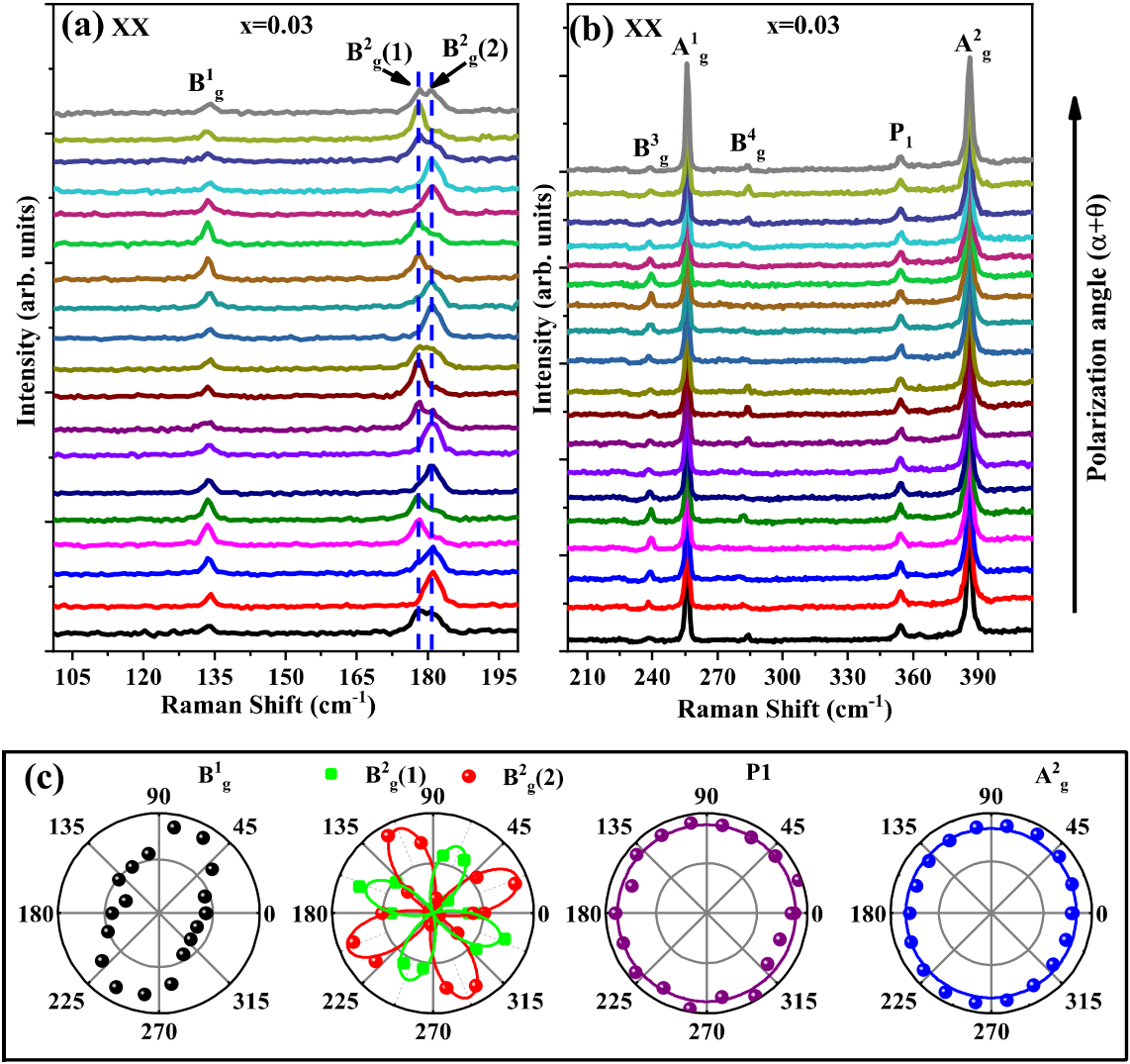

**Figure S4: (a)** and **(b)** Polarization dependent Raman spectra for $x = 0.03$, collected at 3.5 K in the parallel configuration as a function of $\alpha + \theta$, where $\theta$ is the polarization angle varying from the crystallographic *a*-axis and the constant angle $\alpha$ appears due to some deviation in the incident light polarization direction from the crystallographic *a*-axis. **(c)** Intensity polar plot as a function of $\alpha + \theta$.

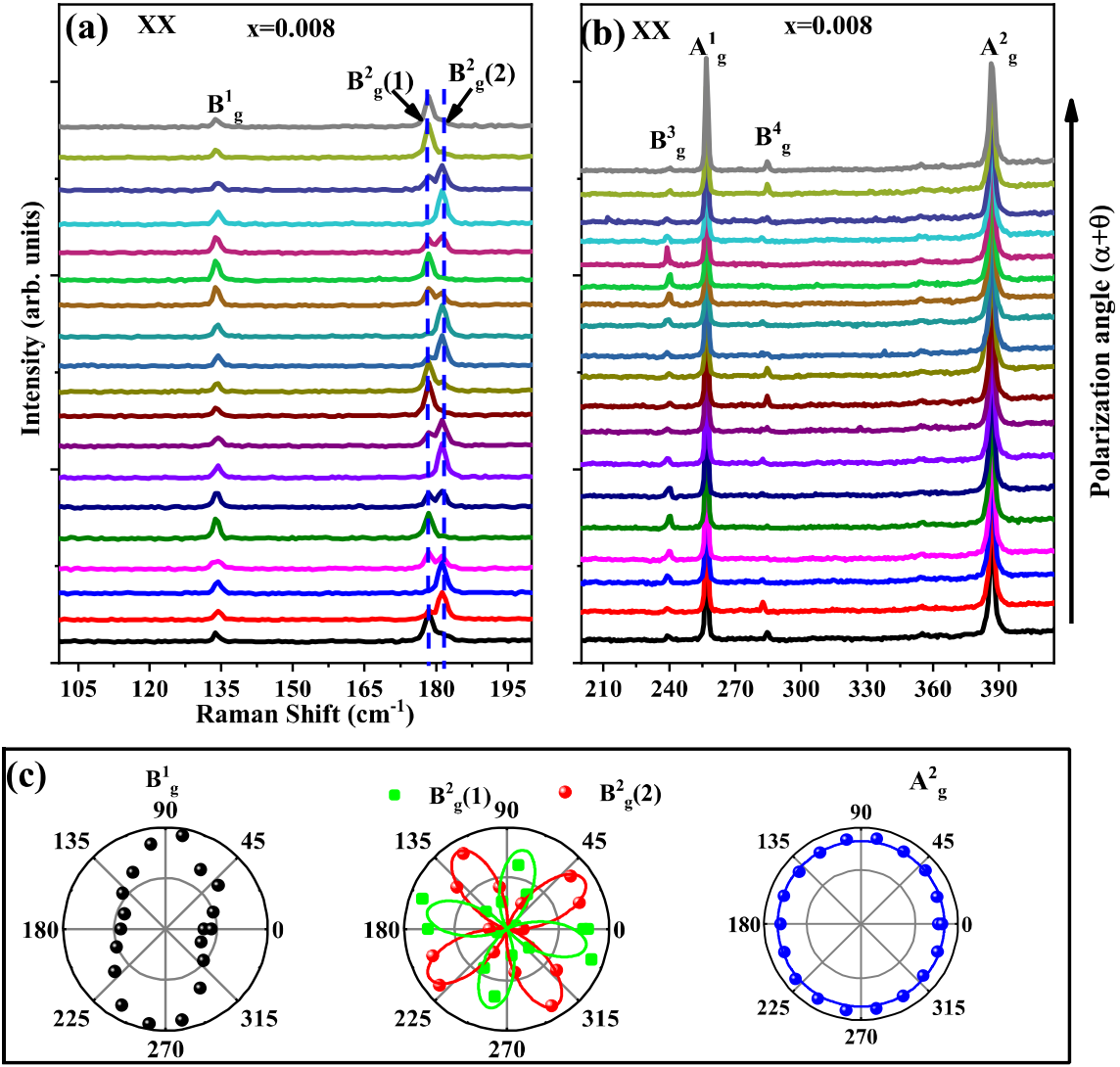

**Figure S5:** **(a)** and **(b)** Polarization dependent Raman spectra for $x = 0.008$, collected at 3.5 K in the parallel configuration as a function of $\alpha + \theta$, where $\theta$ is the polarization angle varying from the crystallographic $a$-axis and $\alpha$ is a constant angle, which is ascribed to some deviation in the incident light polarization direction from the crystallographic $a$-axis. **(c)** Intensity polar plot as a function of $\alpha + \theta$.

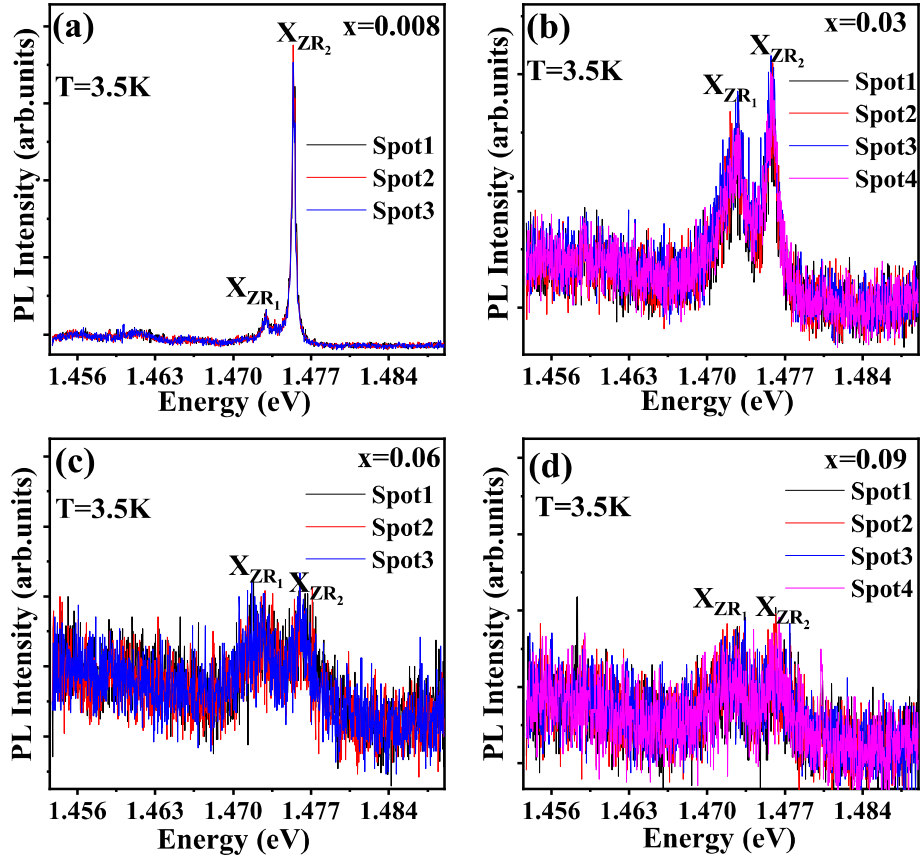

**Figure S6:** PL spectra of NiPS$_{3-x}$Se$_x$ collected from distinct spots.

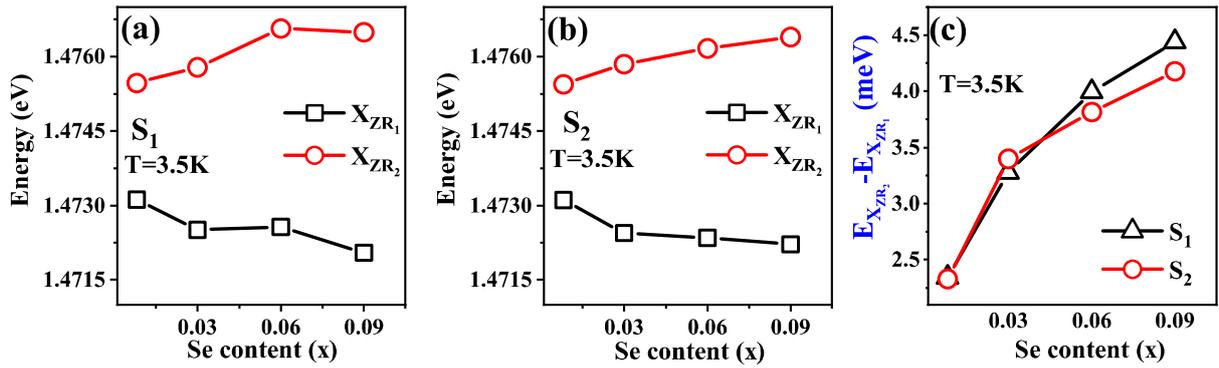

**Figure S7: (a)** and **(b)** Peak energy of $X_{ZR_2}$ and $X_{ZR_1}$ versus Se contents for other two distinct spots (S1 and S2). **(c)** Energy difference between $X_{ZR_2}$ and $X_{ZR_1}$ peak versus Se contents.

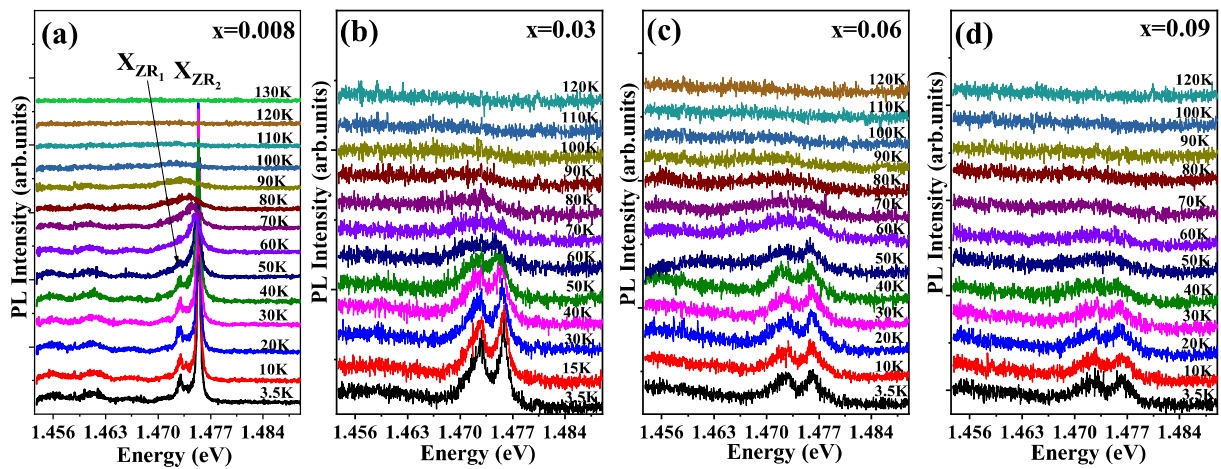

**Figure S8:** Temperature Evolution of the PL spectra for NiPS$_{3-x}$Se$_x$.

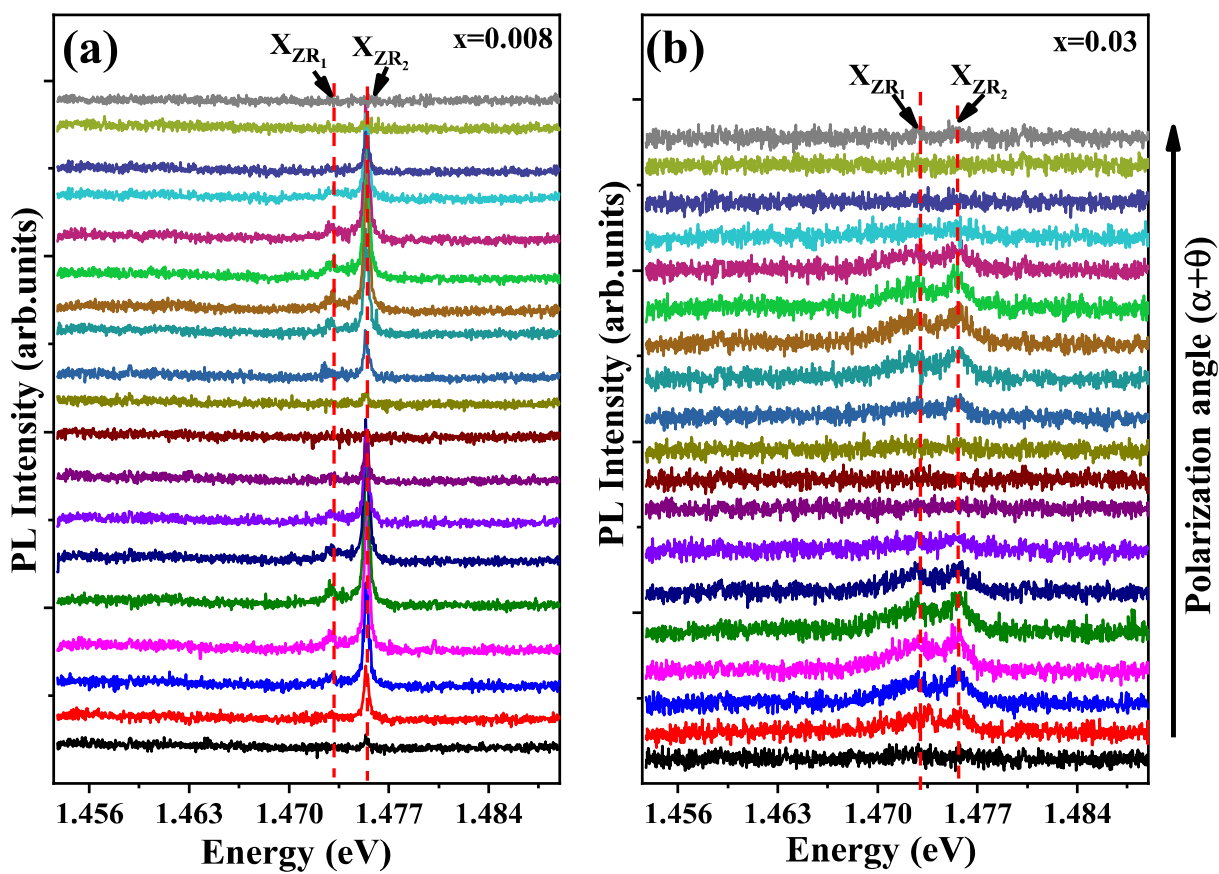

**Figure S9:** Polarization-dependent PL spectra for **(a)** $x = 0.008$ and **(b)** $x = 0.03$, collected at 3.5 K in parallel configuration, as a function of $\alpha + \theta$, where $\theta$ is the polarization angle measured from the crystallographic *a*-axis and $\alpha$ is a constant angle, arising from a deviation in the incident light polarization direction from the crystallographic *a*-axis.

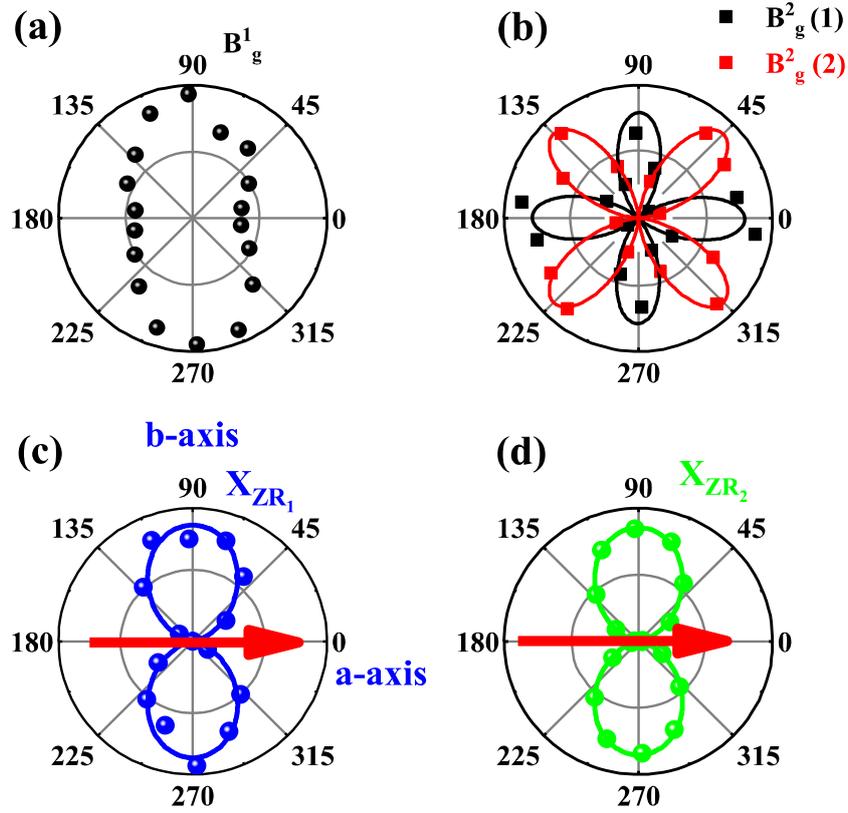

**Figure S10: (a) and (b)** Intensity polar plot for Raman phonons $B_g^1$ and $B_g^2(1,2)$ for $x = 0.008$, collected at 3.5 K in the parallel configuration, after correcting the initial orientation of the crystal axis, respectively. **(c) and (d)** Intensity polar plots of the excitonic peaks $X_{ZR_2}$ and $X_{ZR_1}$ for $x = 0.008$, collected at 3.5 K in the parallel configuration, after the corrections of the initial orientation of the crystal axis, respectively. Red arrows in **(c)** and **(d)** represent the Néel vector which aligns along the *a*-axis. Angular dependence intensity of the $B_g^2(1)$ and $B_g^2(2)$ are described by the expression $I(\theta) = e^2 \cos^2(2\theta)$ and $I(\theta) = e^2 \sin^2(2\theta)$, respectively, while the excitonic peaks $X_{ZR_2}$ and $X_{ZR_1}$ are described by $I(\theta) = I_0 \sin^2(\theta)$.